\documentclass[aps,prl,superscriptaddress,showpacs,twocolumn,floatfix,amsmath,amssymb]{revtex4-1}

\usepackage{hyperref} 
\usepackage[mathlines]{lineno} 

\usepackage{mathrsfs}
\usepackage{bm} 
\usepackage{slashed}
\usepackage{graphicx} 
\usepackage{xcolor}
\usepackage{cancel}

\def\gtorder{\mathrel{\raise.3ex\hbox{$>$}\mkern-14mu
\lower0.6ex\hbox{$\sim$}}}
\def\ltorder{\mathrel{\raise.3ex\hbox{$<$}\mkern-14mu
\lower0.6ex\hbox{$\sim$}}}

\begin{document}

\title{Novel observation of isospin structure of short-range correlations in calcium isotopes}


\newcommand*{\JLAB}{Thomas Jefferson National Accelerator Facility, Newport News, Virginia 23606}
\newcommand*{\TLV}{Tel Aviv University, Tel Aviv 69978, Israel}
\newcommand*{\MIT}{Massachusetts Institute of Technology, Cambridge, Massachusetts 02139}
\newcommand*{\KENT}{Kent State University, Kent, Ohio 44242}
\newcommand*{\DOMINION}{Old Dominion University, Norfolk, Virginia 23529}
\newcommand*{\CALIF}{California State University, Los Angeles, Los Angeles, California 90032}
\newcommand*{\Hampton}{Hampton University, Hampton, Virginia 23668}
\newcommand*{\PENNSYLVANIA}{Pennsylvania State University, State College, Pennsylvania 16801}
\newcommand*{\Paris}{Institut de Physique Nucl\'{e}aire (UMR 8608), CNRS/IN2P3 - Universit\'e Paris-Sud, F-91406 Orsay Cedex, France}
\newcommand*{\Syracuse}{Syracuse University, Syracuse, New York 13244}
\newcommand*{\Kentucky}{University of Kentucky, Lexington, Kentucky 40506}
\newcommand*{\William}{College of William and Mary, Williamsburg, Virginia 23187}
\newcommand*{\Virginia}{University of Virginia, Charlottesville, Virginia 22904}
\newcommand*{\Halifax}{Saint Mary's University, Halifax, Nova Scotia, Canada, B3H 3C3}
\newcommand*{\Glasgow}{University of Glasgow, Glasgow G12 8QQ, Scotland, United Kingdom}
\newcommand*{\Temple}{Temple University, Philadelphia, Pennsylvania 19122}
\newcommand*{\Argonne}{Physics Division, Argonne National Laboratory, Argonne, Illinois 60439}
\newcommand*{\China}{China Institute of Atomic Energy, Beijing, 102413, China}
\newcommand*{\NRCN}{Nuclear Research Center Negev, Beer-Sheva, 84190, Israel}
\newcommand*{\Catania}{Universita di Catania, 95123 Catania, Italy}
\newcommand*{\Dequense}{Duquesne University, Pittsburgh, Pennsylvania 15282}
\newcommand*{\Pittsburgh}{Carnegie Mellon University, Pittsburgh, Pennsylvania 15213}
\newcommand*{\LongwoodUniv}{Longwood University, Farmville, Virginia 23909}
\newcommand*{\Florida}{Florida International University, Miami, Floria 33199}
\newcommand*{\Tallahassee}{Florida State University, Tallahassee, Florida 32306}
\newcommand*{\INFN}{INFN, Sezione Sanit\`{a} and Istituto Superiore di Sanit\`{a}, 00161 Rome, Italy}
\newcommand*{\INFNBari}{INFN, Sezione di Bari and University of Bari, I-70126 Bari, Italy}
\newcommand*{\Ohio}{Ohio University, Athens, Ohio 45701}
\newcommand*{\Tennessee}{University of Tennessee, Knoxville, Tennessee 37996}
\newcommand*{\Kharkov}{Kharkov Institute of Physics and Technology, Kharkov 61108, Ukraine}
\newcommand*{\LOSALAMOS}{Los Alamos National Laboratory, Los Alamos, New Mexico 87545}
\newcommand*{\Duke}{Duke University, Durham, North Carolina 27708}
\newcommand*{\Texas}{University of Texas, Houston, Texas 77030}
\newcommand*{\Seoul}{Seoul National University, Seoul 151-742, Korea}
\newcommand*{\Indiana}{Indiana University, Bloomington, Indiana 47405}
\newcommand*{\Hampshire}{University of New Hampshire, Durham, New Hampshire 03824}
\newcommand*{\Blacksburg}{Virginia Polytechnic Inst. and State Univ., Blacksburg, Virginia 24061}
\newcommand*{\France}{Universit\'e Blaise Pascal/IN2P3, F-63177 Aubi\`ere, France}
\newcommand*{\Mississippi}{Mississippi State University, Starkville, Mississippi 39762}
\newcommand*{\Austin}{The University of Texas at Austin, Austin, Texas 78712}
\newcommand*{\Norfolk}{Norfolk State University, Norfolk, Virginia 23504}
\newcommand*{\Lanzhou}{Lanzhou University, Lanzhou 730030, China}
\newcommand*{\Hebrew}{Racah Institute of Physics, Hebrew University of Jerusalem, Jerusalem 91905, Israel}
\newcommand*{\Rutgers}{Rutgers, The State University of New Jersey, Piscataway, New Jersey 08855}
\newcommand*{\Yerevan}{Yerevan Physics Institute, Yerevan 375036, Armenia}
\newcommand*{\Ljubljana}{Faculty of Mathematics and Physics, University of Ljubljana, Ljubljana 1000, Slovenia}
\newcommand*{\Michigan}{Northern Michigan University, Marquette, MI 49855}
\newcommand*{\Hefei}{University of Science and Technology, Hefei 230052, China}	
\newcommand*{\Jozef}{Jozef Stefan Institute, Ljubljana 1001, Slovenia}
\newcommand*{\Ecole}{CEA Saclay, F-91191 Gif-sur-Yvette, France}
\newcommand*{\Massachusetts}{University of Massachusetts, Amherst, Massachusetts 01006}
\newcommand*{\TAMUK}{Texas A\&M University - Kingsville, Kingsville, Texas, 78363}

\author{D. Nguyen}
\email[contact email: ]{dien@mit.edu}
\affiliation{\Virginia}
\affiliation{\MIT}
\author{Z. Ye}
\email[contact email: ]{zhihong@jlab.org}
\thanks{present address: Canon Medical Research USA}
\affiliation{\Virginia}
\affiliation{\Argonne}
\author{P. Aguilera}
\affiliation{\Paris}
\author{Z. Ahmed}
\affiliation{\Syracuse}
\author{H. Albataineh}
\affiliation{\TAMUK}
\author{K. Allada}
\affiliation{\JLAB}
\author{B. Anderson}
\affiliation{\KENT}
\author{D. Anez}
\affiliation{\Halifax}	
\author{K.~Aniol}
\affiliation{\CALIF}
\author{J. Annand}
\affiliation{\Glasgow}
% ...

\collaboration{The Jefferson Lab Hall A Collaboration}


\date{\today}

\begin{abstract}
Short Range Correlations (SRCs) have been identified as being responsible for the high momentum tail of the nucleon momentum distribution, $n(k)$. Hard, short-range interactions of nucleon pairs generate the high momentum tail and imprint a universal character on $n(k)$ for all nuclei at large momentum. Triple coincidence experiments have shown a strong dominance of $np$ pairs, but these measurements involve large final state interactions. This paper presents the results from Jefferson Lab experiment E08014 which measured inclusive electron scattering cross-section from Ca isotopes. By comparing the inclusive cross section from $^{48}$Ca to $^{40}$Ca in a kinematic region dominated by SRCs we provide a new way to study the isospin structure of SRCs.
\end{abstract}

\maketitle

\section{Introduction}
The na\"ive nuclear shell model has guided our understanding of nuclear properties for 60 years and it is still appealing as a predictive and illustrative nuclear model. This model, with nucleons moving in an average mean-field generated by the other nucleons in the nucleus, provides a quantitative account of a large body of nuclear properties. These include shell closures (``magic numbers''), the foundation of which is the appearance of gaps in the spectrum of single-particle energies~\cite{caurier05}.

The shell model is not without certain deficits which arise from what are generally called correlations - effects that are beyond mean field theories such as long-range correlations associated with collective phenomena: giant resonances, vibrations and rotations.
In addition, electron-nucleus and nucleon-nucleus scattering experiments have unambiguously shown large deviations from the shell model predictions, arising from the occurrence of strong short-range nucleon-nucleon correlations. These two-nucleon SRCs (2N-SRC) move particles from the shell model states to large excitation energies and generate a high-momentum tail in the single particle momentum distribution. Consequently, over a large range in $A$ the number of protons found in the valence shells orbitals is significantly less than expected, typically 60\%--70\% of the predicted shell model occupancy~\cite{kelly96, Lapikas1993297}.

Inclusive experiments are able to isolate the 2N-SRC through selective kinematics: working at large momentum transfer ($Q^2 \ge 1.5 $~(GeV/$c$)$^2$) and small energy transfer ($\nu \le \frac{Q^2}{2m}$), corresponding to $x = \frac{Q^2}{2m\nu} > 1$, where $m$ is the mass of the proton. In these kinematics, inelastic scattering is minimized and quasielastic scattering requires that the struck nucleon have a non-zero initial momentum, as scattering at $x>1$ is kinematically forbidden for a stationary nucleon~\cite{frankfurt81, sargsian03, arrington12a}. By selecting sufficiently large $x$ and $Q^2$, the minimum initial nucleon momentum can be set above the Fermi momentum, dramatically suppressing the contribution from mean-field nucleons and isolating scattering from 2N-SRCs.  It was through inclusive experiments~\cite{frankfurt93, egiyan03, PhysRevLett.96.082501, fomin12} that 2N-SRCs were first revealed by the appearance of predicted plateaus~\cite{frankfurt81} in the $A$/$^2$H per-nucleon cross section ratio of nuclei to the deuteron. The height of the plateau is related to probability of finding a 2N-SRC in nucleus $A$, relative to the deuteron, indicating that $\sim$20\% of the protons and neutrons in medium-to-heavy mass nuclei have momenta greater than the Fermi momentum $k_F \simeq 250$~MeV/$c$~\cite{PhysRevLett.96.082501}. The bulk of these nucleons do not arise in a shell model description as they are the result of brief short-range interaction among pairs of nucleons giving rise to large relative momenta and modest center-of-mass motion, $k_{CM} < k_F$~\cite{frankfurt81}. 

The isospin dependence of 2N-SRCs has been determined via $A(p,p'pN)$~\cite{tang03,Piasetzky:2006ai} and $A(e,e'pN)$~\cite{Shneor:2007tu, Subedi:2008zz, Korover:2014dma} reactions in which the scattered particle (either a proton or an electron) is measured in coincidence with a high-momentum proton. The struck proton's initial momentum, reconstructed assuming plane wave scattering, is approximately opposite that of the second high-momentum nucleon.
These measurements exhibit a dominance of $np$ pairs over $pp$ pairs for initial nucleon momenta of 300-600~MeV/$c$ which has been traced to  the tensor part of the $NN$ interaction~\cite{PhysRevC.71.044615, schiavilla07, alvioli08, wiringa08, wiringa14}. These triple-coincidence experiments are sensitive to the isospin structure of the SRC through direct measurement of the final state nucleons. Because the signature of large back-to-back momenta is also consistent with striking a low-momentum nucleon which rescatters, there are large contributions from final state interactions (including charge exchange) that need to be accounted for in comparing $pp$ and $np$ pairs~\cite{Shneor:2007tu,Subedi:2008zz,Korover:2014dma}. Isospin dependence has never been established in inclusive scattering, $A(e,e')$ until now.

We present here new $A(e,e')$ measurements performed as part of Jefferson Lab experiment E08-014~\cite{e08014_pr}. Initial results on the search for three-nucleon SRCs in helium isotopes were published in Ref.~\cite{ye2019}. The present work focuses on a measurement of the isospin dependence of 2N-SRCs in the cross section ratio of scattering from $^{48}$Ca and $^{40}$Ca. The excess neutrons in $^{48}$Ca change the relative ratio of potential $pp$, $np$, and $nn$ pairs, but the impact on the cross section depends on whether the generation of high-momentum pairs is isospin independent or $np$ dominated. This can be illustrated by making a very simple estimate. As a starting point, we take the fraction of nucleons in SRCs to be identical for these two targets, based on the observation of an A-independent value of $a_2$, the $A$/$^2$H cross section ratio for $1.5<x<2$, for heavy nuclei~\cite{frankfurt93, fomin12}. In the case of isospin-independent SRCs, protons and neutrons will have the same probability of appearing at momenta above $k_F$, giving a cross section ratio of $\frac{\sigma_{^{48}\text{Ca}}/48}{\sigma_{^{40}\text{Ca}}/40} = \frac{(20\sigma_{ep} + 28\sigma_{en})/48}{(20\sigma_{ep} + 20\sigma_{en})/40} \approx 0.93$ taking $\sigma_{ep} / \sigma_{en} \approx 2.5$, corresponding at the kinematics of this experiment. If SRCs are dominated by $np$ pairs, the cross section ratio would be unity for isoscalar nuclei and slightly lower for non-isoscalar nuclei~\cite{vanhalst11, vanhalst12, Ryckebusch:2018rct, Ryckebusch:2019oya}. A more detailed cross section model is presented later and used to interpret the data.

\section{Experimental details}
Jefferson Lab experiment E08-014~\cite{e08014_pr} ran during the Spring of 2011. A 3.356~GeV continuous wave electron beam was directed onto a variety of targets, including $^2$H, $^3$He, $^4$He, $^{12}$C and targets of natural calcium (mainly $^{40}$Ca) and an enriched target of 90.04\% $^{48}$Ca (referred to as the $^{40}$Ca and $^{48}$Ca targets, respectively). The scattered electrons were detected at angles of $\theta$=21$^{\circ}$, 23$^{\circ}$, 25$^{\circ}$, and 28$^{\circ}$, though no calcium data were taken at 28$^{\circ}$. The data presented here cover a kinematic region of $1.3 < Q^2 < 1.9$~(GeV/$c$)$^2$ and $1<x<3$.

The inclusive scattered electrons were detected using two nearly identical left and right high resolution spectrometers (LHRS and RHRS). The spectrometers consisted of a set of three quadrupole magnets and a dipole magnet which transported the events from the target to the detector plane. Each spectrometer was equipped with a detector package consisting of two vertical drift chambers (VDC) for tracking information~\cite{Fissum:2001st}, a Gas Cerenkov counter~\cite{Iodice:1998ft} and two layers of lead glass calorimeters for particle identification (PID), and two scintillator counter planes for triggering~\cite{halla_nim}.

The accumulated charge for each experimental run was measured by beam current monitors with an uncertainty of 0.5\%, based on the difference in the calculated charge using two sets of beam calibration parameters. The dead-time due to the inability of the data acquisition system to accept new triggers while processing an earlier event was corrected for each run using the trigger scaler information. The main trigger for data collection required a coincidence of signals from two scintillator planes and the Cerenkov, which had a local inefficiency region. A sample of events were taken with a second trigger which did not require the Cerenkov signal, allowing for continuous measurement of the inefficiency. The correction was applied to the measured yield in each $x$ bin for every kinematic setting. Pions were rejected (with negligible remaining pion contamination) by applying additional cuts on both the Cerenkov counter and the lead glass calorimeter with efficiencies of 99.5\% and 99.6\% respectively, with a tracking efficiency of 98.5\%. Detailed descriptions of the experimental setup and data analysis can be found in ~\cite{dien_thesis, zye_thesis}.

The position and angle of the scattered electrons are measured at the VDCs. These are used to reconstruct the angle and momentum of the scattered electron using reconstruction matrices which parameterize the transformation between electron trajectory at the target and the VDCs. These were determined by a fit to special data runs where the particle angle or momentum were determined by taking data in over-constrained kinematics (e.g. elastic scattering) to define the momentum or using a collimator with small holes at the entrance to the first HRS magnet~\cite{halla_nim} to define the scattering angle. For the left HRS, the reconstruction matrix was fitted to data taken from the previous experiment~\cite{e1207108} which had the same spectrometer and magnet settings as this experiment. The magnet tune for the RHRS had to be modified because the third quadrupole couldn't run at the desired field, and lack of a complete set of calibration data for to fit led to a reduced resolution in the RHRS. The reduced resolution affected the extraction of the cross section at large $x$ values where the cross section falls extremely rapidly, requiring a larger correction. Because the RHRS was typically taking data in the same kinematics as the LHRS, we used only the data from the LHRS except for the 21$^\circ$ data, where the largest $x$ values were measured only in the RHRS. For this setting we included the ratios from the RHRS, as the smearing has a negligible impact on the cross section ratios in the region where the ratio is flat. 

The yield for the experiment was simulated using a detailed model of the HRS optics and acceptance, with events generated uniformly and weighted by a radiative cross section model~\cite{zye_thesis, zye_xemc}. The model used a $y$-scaling fit~\cite{west75, day90} for quasi-elastic cross section (initially based on previous data, and iteratively updated to match the extracted cross sections from this experiment) and a global fit~\cite{bosted12} for the inelastic contribution.  The Born cross section was extracted by taking the model cross section and correcting it by the ratio of measured to simulated yield. Comparing the results extracted with the final model and the model before being adjusted to match our data indicated a model uncertainty of 0.5\% in both the absolute cross sections and the target ratios. 

The cross section ratio obtained from the enriched and natural calcium targets was then corrected to yield $^{48}$Ca/$^{40}$Ca ratio, based on the isotopic analysis of the targets. No correction was applied to the natural calcium, which was over 99.9\% $^{40}$Ca. The enriched calcium target was 90.04\% $^{48}$Ca and 9.96\% $^{40}$Ca, by number of atoms. Using the measured $^{40}$Ca cross section, we correct for the $^{40}$Ca contribution in the enriched target to extract the $^{48}$Ca cross section. The correction is typically 0.5-1.5\%.


\section{Results}

\begin{figure}[htb]
\centering
\includegraphics[width=0.5\textwidth]{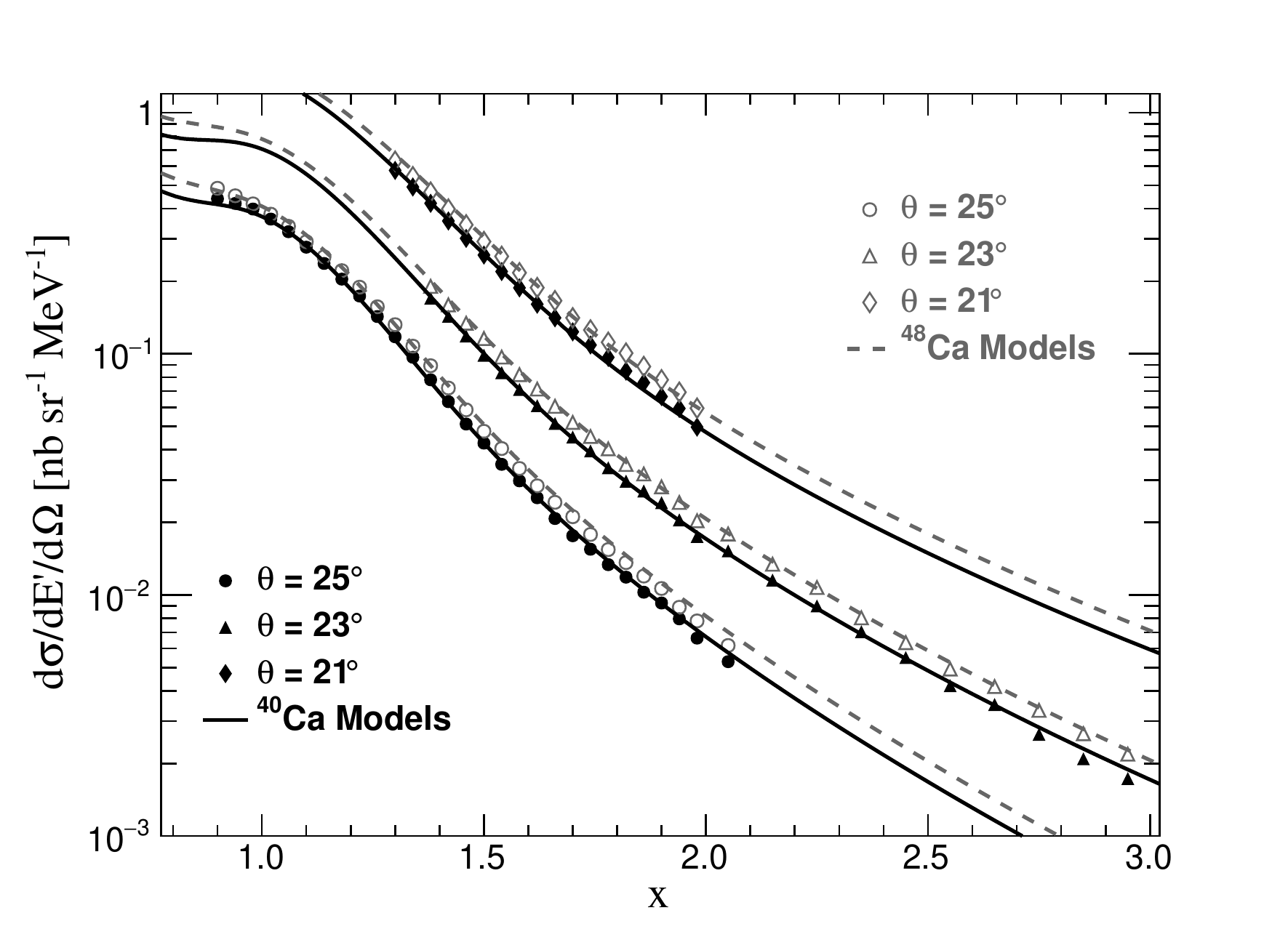}
\caption{$^{48}$Ca and $^{40}$Ca cross sections for three different angle settings, along with the cross section model used in the analysis. Uncertainties shown include statistical and point-to-point systematic uncertainties; an additional normalization uncertainty of 2.7\% for $^{40}$Ca and 3.0\% for $^{48}$Ca is not shown.}
\label{fig:ca_xs}
\end{figure}

The measured cross sections are presented in Figure~\ref{fig:ca_xs}. For the cross sections, the point-to-point systematic uncertainty is estimated to be 1.9\%, with dominant contribution coming from the acceptance (1.5\%), radiative corrections (1\%), and the model dependence of the cross section extraction (0.5\%).  In addition, there is an overall normalization uncertainty of 2.7\%, coming mainly from the acceptance (2\%), radiative correction (1\%), and target thickness (1\%). These are the uncertainties for the $^{40}$Ca target. The dilution correction used to extract the $^{48}$Ca cross section increases these, giving 2.1\% point-to-point and 3.0\% normalization uncertainties. 

The point-to-point uncertainty due to the acceptance was determined by systematically selecting five different sets of acceptance cuts. The uncertainty was taken to be the variation in the extracted cross sections corresponding to these cuts, which was consistent with previous estimates of the HRS acceptance.
The cross section model in this analysis was iterated three times. The variation in the extracted cross sections using different iterated models was assigned as the point-to-point uncertainty for the model dependence. Radiative corrections were applied using the prescription described in~\cite{PhysRevD.12.1884}, with uncertainties that account for limitations of the procedure and uncertainty in the energy loss and radiation length of the target material.

\begin{figure}[htb]
\centering
\includegraphics[width=0.52\textwidth]{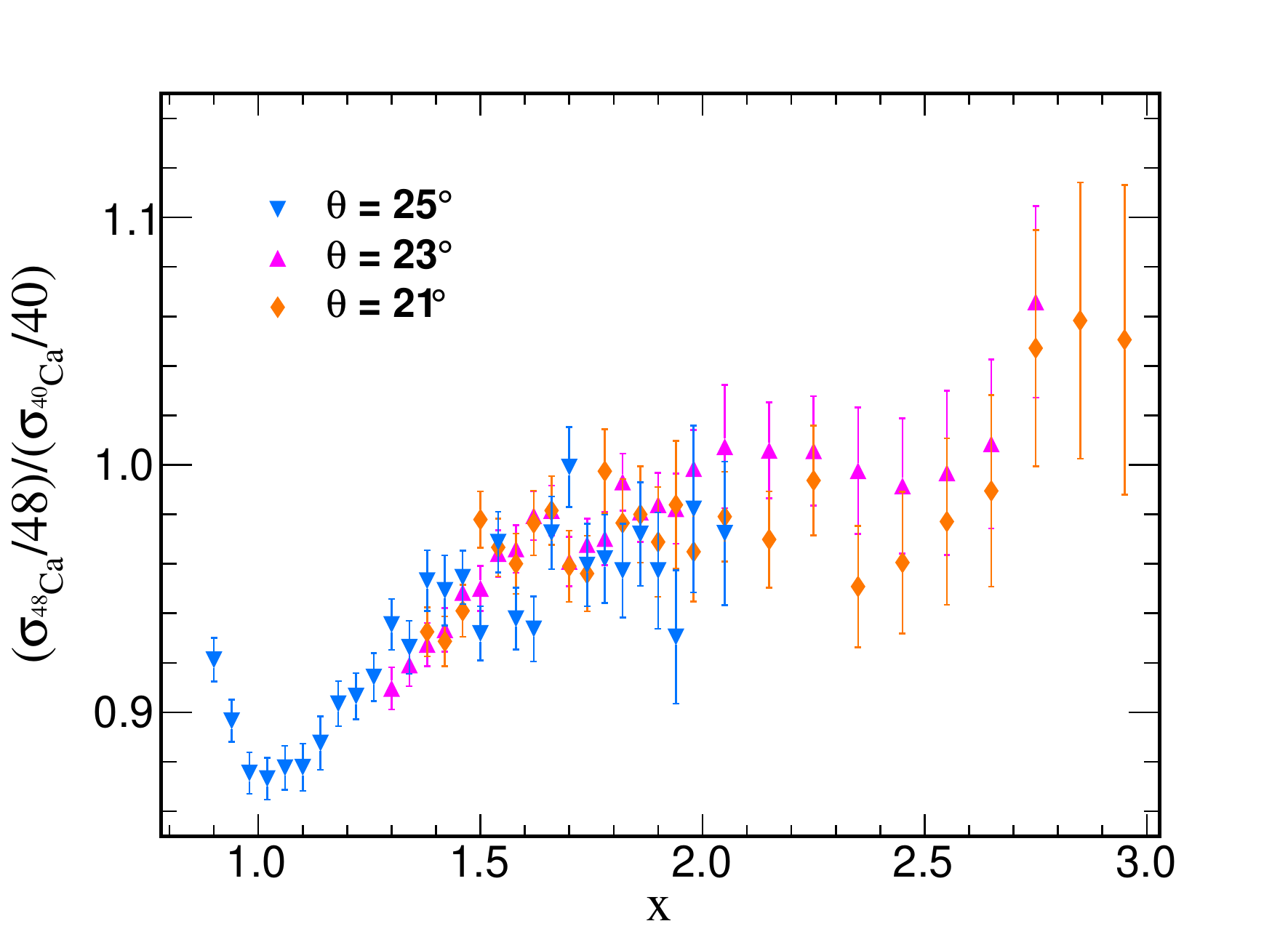}
\caption{Ratio of the cross section per nucleon for $^{48}$Ca and $^{40}$Ca for three scattering angles. Uncertainties shown include statistical and point-to-point systematic uncertainties; an additional normalization uncertainty of 1\% is not shown.}
\label{fig:ca_ratio_angles}
\end{figure}

\begin{figure}[htb]
\centering
\includegraphics[width=0.52\textwidth]{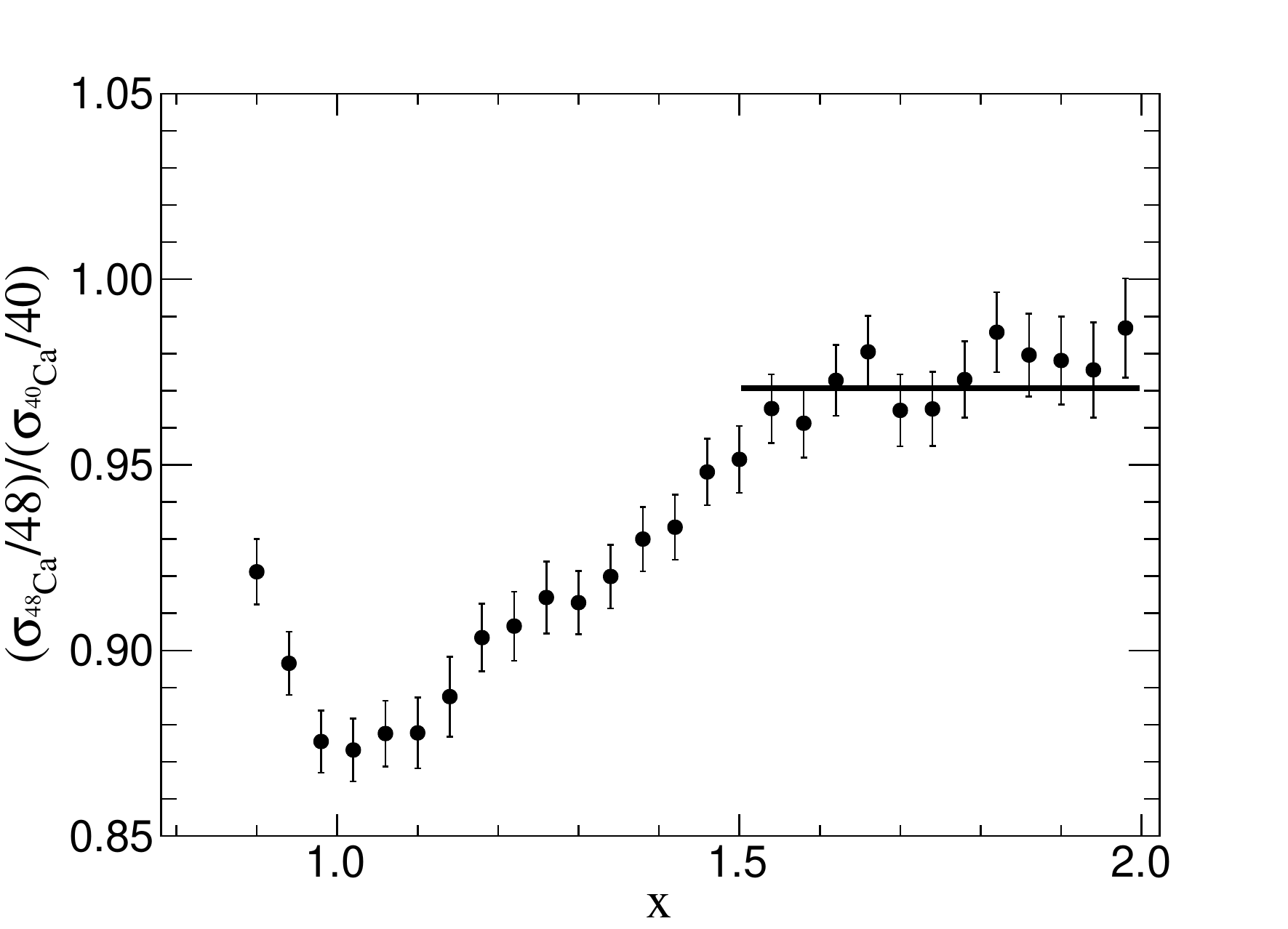}
\caption{Ratio of the cross section per nucleon for $^{48}$Ca and $^{40}$Ca combining all three data sets. A 1\% normalization uncertainty is not shown. The line indicates the fit for the cross section ratio in the SRC region}
\label{fig:ca_ratio} 
\end{figure}

The per nucleon cross-section ratio of $^{48}$Ca to $^{40}$Ca is presented in Figure~\ref{fig:ca_ratio_angles} for each of the three scattering angles and in Figure~\ref{fig:ca_ratio} after combining of the data sets. Because the cross section and experimental conditions are very similar for the two targets, many of the uncertainties in the cross sections cancel or are reduced in the ratio. The systematic uncertainty on the ratios is 0.9\%, dominated by the model dependence in the extraction (0.5\%), measurement of the beam charge (0.5\%) and the radiative correction (0.5\%). An additional 1\% normalization uncertainty, associated with the uncertainty in the relative target thicknesses, is not shown. For Fig.~\ref{fig:ca_ratio}, we combined the statistics of the individual sets (for 3 angles) and then apply the 0.9\% point-to-point uncertainty (and 1\% normalization uncertainty) to the result. 

Note that the rise from $x=1$ to $x=1.6$ looks slightly different than it does for the ratios to $^2$H~\cite{frankfurt93, fomin12}. This is expected as the shape in the $A$/$^2$H ratios is driven by the deuterium cross section, which is narrowly peaked at and roughly symmetric about $x=1$. The line in Fig.~\ref{fig:ca_ratio} indicates the value of $R_{\mbox{SRC}}$, the average in the plateau region: $1.5 < x < 2$. The fit gives $R_{\mbox{SRC}}$=0.971(3)(6)(10)=0.971(12) where the error contributions come from the point-to-point uncertainties, the cut dependence of the extracted $R_{\mbox{SRC}}$, and the normalization uncertainty of the ratios. The cut dependence is taken to be the RMS scatter of $R_{\mbox{SRC}}$ values fit separately to the three scattering angles for three different minimum $x$ values, $x_{min}=1.5, 1.6$ and 1.7.

The observed value of $R_{\mbox{SRC}}=0.971(12)$ is more than three sigma above the prediction for isospin independence ($R_{\mbox{SRC}} = 0.930$ for these kinematics). So while inclusive scattering cannot isolate contributions from protons and neutrons, comparing calcium isotopes with significantly different $N$/$Z$ values is sensitive enough to provide evidence for an enhancement of $np$ pairs over $pp$ and $nn$ pairs.

\section{Discussion}
To quantitatively interpret this ratio in terms of relative $np$, $pp$, and $nn$ SRC contributions, and to compare these results to observable from previous measurements, we use a simple model to estimate the inclusive, exclusive, and two-nucleon knockout ratios in terms of a few parameters. We take the number of 2N-SRCs to be a product of the number of total pairs, the probability for any two nucleons to be close enough together to interact via the short-range $NN$ interaction ($f_{sr}$), and the probability that the $NN$ interaction generates a high-momentum pair ($p_{NN}$). The total number of $np$, $pp$, and $nn$ pairs are $NZ$, $Z(Z-1)/2$, and $N(N-1)/2$, respectively. The fraction of nucleons at short distance, $f_{sr}$, depends on the nucleus and is assumed to be identical for $nn$, $np$, and $pp$ pairs. The probability that these nucleons generate high momentum pairs, $p_{np}$ and $p_{pp} = p_{nn}$, depends on the momentum range of the initial nucleons, $\Delta P_i$, defined by the experiment for coincidence measurements or by the kinematics in inclusive scattering. Given this, we can express the number of $np$ and $pp$ SRCs as:

\begin{align}
\label{eq:NN_pairs2}
N_{np} & = N Z \cdot f_{sr}(A) \cdot p_{np}(\Delta P_i) \\
N_{pp} & = Z(Z-1)/2 \cdot f_{sr}(A) \cdot p_{pp}(\Delta P_i)
\end{align}

While $p_{np}$ and $p_{pp}$ may depend strongly on $\Delta P_i$, we assume that their ratio has a much weaker dependence, as observed in Ref.~\cite{Korover:2014dma}, and so their ratio extracted from different measurements should be comparable. This leaves only $f_{sr}(A)$ as an unknown. In comparing different observables on the same nucleus, e.g. taking the ratio of $A(e,e'pp)$ to $A(e,e'pn)$, $f_{sr}(A)$ cancels out. In the limit of large nuclei, any given nucleon will be sensitive to short-range interactions with nucleons in some fixed volume, while the number of nucleons grows with $A$, suggesting that $f_{sr}(A)$ should scale as $1/A$. With this assumption, our model produces a constant value of $a_2$ for heavy isoscalar nuclei, consistent with the observation of approximate saturation~\cite{arrington12b}. Note that the result $f_{sr}(A) \propto 1/A$ is derived assuming np dominance. Under the isospin-independent assumption, $f_{sr}(A)$ must scale as $1/(A-1)$. For the comparison of $^{48}$Ca to $^{40}$Ca, the difference between these cases is less than 0.5\%.

Within this model, we can calculate the number of $pp$, $np$, and $nn$ SRCs in the $^{40}$Ca and $^{48}$Ca targets. Each SRC pair contributes to the incusive cross section in the SRC-dominated region based on the $e$-$N$ elastic cross section for the two nucleons. In cross section ratio, only the $A$ dependence of $f_{sr}(A)$ remains and the ratio depends only on the $A$ dependence of $f_{sr}(A)$ which is taken to scale as $1/A$. The cross section ratio thus depends only on the ratio of electron-proton to electron-neutron elastic scattering, and on the ratio $p_{np}$/$p_{pp}$, the enhancement factor of $np$ pairs to high momentum relative to $pp$ (and $nn$) pairs. The average value of $\sigma_{ep}/\sigma_{en}$ is 2.55-2.60 for these kinematics, this model predicts the cross section ratio to be 0.930 for isospin independence, and 0.972 for $np$ dominance.

The observed cross section ratio is significantly above the prediction for isospin-independent pairing. Taking into account its uncertainty, we find that $p_{np}/p_{pp}>2.9$ at the 95\% confidence level, and $p_{np}/p_{pp}>1.6$ at the 99\% confidence level, demonstrating $np$ dominance using the isospin structure of the target, rather than the detected nucleons, to study the isospin structure.

Our prediction for the isospin-independent ratio neglects the difference between the size of the proton and neutron distributions in $^{48}$Ca. Based on the estimated charge radius~\cite{charge_radii} and a neutron skin~\cite{neutron_skin}, the RMS radii of the proton and neutron distributions are 3.5~fm and 3.7~fm, respectively. Relative to our model, which assumes a uniform radius of 3.6~fm, the proton (neutron) distribution is roughly 8.5\% smaller (larger) in volume which will increase proton pairing and decrease neutron pairing by similar amounts. Because the $pp$ and $nn$ pairs contribute nearly equally to the $x>1$ cross section in $^{48}$Ca, the net impact on the cross section is a very small ($<$0.5\%) increase in the $^{48}$Ca cross section. The contribution from np pairs, which dominate the cross section, will be decreased due to the reduced overlap between the proton and neutron distributions, providing a modest reduction to the $^{48}$Ca cross section. The net effect should be a decrease in the prediction of $R_{\mbox{SRC}}=0.930$ for the isospin-independent model, although the size of the effect is difficult to estimate precisely. However, taking $R_{\mbox{SRC}}=0.930$ as an upper limit, the indications for np-dominance seen in the data will be at least as significant as estimated above.  

The ratio $p_{np}/p_{pp}$ cannot be directly compared to the enhancement factor of $\sim$10 obtained in triple coincidence experiments~\cite{Subedi:2008zz,Korover:2014dma}, as it removes the contribution from simple pair counting. For example, $^4$He has four $np$ pairs and only one $pp$ pair, and thus one would expect $np$ pairs to dominate, even if the generation of high-momentum pairs had no isospin dependence. Using our simple model we can extract $p_{np}/p_{pp}$ from other measurements, $A(e,e'pp)/A(e,e'pn)$ or $A(e,e'p)/A(e,e'n)$, allowing for a more direct comparison. As noted before, $p_{np}$ and $p_{pp}$ depend on the momentum of the struck nucleon in the initial-state SRC, while for the inclusive case, they correspond to an average over the momentum range probed in the scattering which depends on $Q^2$ and the $x$ range of the data. Because of this, the extracted enhancement factor for inclusive scattering corresponds to a range of momenta that should be similar, but not identical, to the momentum range selected in the coincidence knockout reactions. 

Writing out the ratio of $A(e,e'pp)/A(e,e'pn)$ in terms of $p_{np}/p_{pp}$ allows us to take the observed ratios and extract the $np$ enhancement factor. For $^4$He~\cite{Korover:2014dma}, the $pp/np$ fraction is (5.5$\pm$3)\%, giving a one-sigma range of $2.9< p_{np}/p_{pp} <10$. For $^{12}$C, the $pp/np$ fraction is (5.6$\pm$1.8)\% leading to a one-sigma range of $5.6< p_{np}/p_{pp} <11$. The full expressions are provided in the supplementary material~\cite{supp}. The triple-coincidence measurements quote their results in $P_m$ (missing momentum) bins, which represent the reconstructed initial momentum of the struck nucleon~\cite{Korover:2014dma}. We take the lowest $P_m$ bins from the triple-coincidence measurements, covering momenta from 300-600 MeV/$c$, to more closely match the main contributions to the inclusive measurement. As noted above, these values are not exactly equivalent to the values extracted from the inclusive scattering, but they paint a consistent picture of significant $np$ dominance in SRCs over a range of light and heavy nuclei.

\section{Conclusions}
In conclusion, the per nucleon cross section ratio of $^{48}$Ca/$^{40}$Ca is consistent with significant $np$ dominance in the creation of SRCs. It shows an enhancement of $np$ pairs over $pp$ pairs at more than the three sigma level.

This data provides the first evidence of $np$ dominance from inclusive scattering, making use of the isospin structure of the target rather than the final $NN$ pair. This approach avoids the significant corrections required to interpret triple-coincidence measurements, but does not provide a quantitative measure of the enhancement factor because of the small difference between isospin-independent and $np$-dominance assumptions. A recent experiment measured the inclusive ratio for scattering from $^3$H and $^3$He, which is significantly more sensitive~\cite{e1211112}. The  $^3$H/$^3$He cross section ratio is approximately 0.75 for isospin independence and 1 for $np$ dominance, giving almost an order of magnitude more sensitivity than the $^{48}$Ca/$^{40}$Ca ratio, without having to make any assumption about the $A$ dependence of $f_{sr}(A)$ in comparing the two nuclei. A measurement of this inclusive cross section ratio with comparable uncertainties may provide the best quantitative measurement of the enhancement of $np$ pairs at high momentum.

\section{acknowledgments}

We acknowledge the outstanding support from the Hall A technical staff and the JLab target group. This work was supported in part by the Department of Energy's Office of Science, Office of Nuclear Physics, under contracts DE-AC02-06CH11357 and DE-FG02-96ER40950, and the National Science Foundation, and under DOE contract DE-AC05-06OR23177, under which JSA, LLC operates JLab. The $^{48}$Ca isotope used in this research was supplied by the Isotope Program within the Office of Nuclear Physics in the Department of the Energy's Office of Science. Experiment E08-014 was developed by Patricia Solvignon whose passing is still mourned by our community.



%
\end{document}